\documentclass{article}
\usepackage{amssymb,amsmath}
\usepackage{epsfig}
\usepackage{subfigure}
\newtheorem{rem}{Remark}[section]
\newcommand{\br}{\begin{rem}}
\newcommand{\er}{\end{rem}}
\newtheorem{ex}{Example}[section]
\newcommand{\bex}{\begin{ex}}
\newcommand{\eex}{\end{ex}}
\newtheorem{Def}{Definition}[section]
\newcommand{\bd}{\begin{Def}}
\newcommand{\ed}{\end{Def}}
\newtheorem{theorem}{Theorem}[section]
\newcommand{\bt}{\begin{theorem}}
\newcommand{\et}{\end{theorem}}
\newtheorem{lemma}{Lemma}[section]
\newcommand{\bl}{\begin{lemma}}
\newcommand{\el}{\end{lemma}}
\newcommand{\be}{\begin{equation}}
\newcommand{\ee}{\end{equation}}
\newcommand{\bea}{\begin{eqnarray}}
\newcommand{\eea}{\end{eqnarray}}
\newcommand{\pa}{\partial}
\newcommand{\nn}{\nonumber}
\newcommand{\adots}{\mathinner{\mkern2mu\raise1pt\hbox{.}\mkern2mu
\raise4pt\hbox{.}\mkern2mu\raise7pt\hbox{.}\mkern1mu}}

\title{Classical and Quantum Super-integrability:\\  From Lissajous Figures to Exact Solvability}
\author{Allan Fordy, School of Mathematics, \\
University of Leeds\\
e-mail a.p.fordy@leeds.ac.uk}

\begin{document}

\maketitle

\begin{abstract}
The first part of this paper explains what super-integrability is and how it differs in the classical and quantum cases.  This is illustrated with an elementary example of the resonant harmonic oscillator.

For Hamiltonians in ``natural form'', the kinetic energy has geometric origins and, in the flat and constant curvature cases, the large isometry group plays a vital role.  We explain how to use the corresponding first integrals to build separable and super-integrable systems.  We also show how to use the automorphisms of the symmetry algebra to help build the Poisson relations of the corresponding non-Abelian Poisson algebra.

Finally, we take both the classical and quantum Zernike system, recently discussed by Pogosyan, et al, and show how the algebraic structure of its super-integrability can be understood in this framework.
\end{abstract}



\section{Introduction}

We start this paper with a brief reminder of the definitions of complete and super-integrability.

A Hamiltonian system of $n$ degrees of freedom, Hamiltonian $H$, is said to be {\em completely integrable in the Liouville sense} if we have $n$ independent functions $I_n$, which are {\em in involution} (mutually Poisson commuting), with $H$ being a function of these and typically just one of them. Whilst $n$ is the maximal number of independent functions which can be {\em in involution}, it is possible to have further integrals of the Hamiltonian $H$, which necessarily generate a non-Abelian algebra of integrals of $H$.  The maximal number of additional {\em independent} integrals is $n-1$, since the ``level surface'' of $2n-1$ integrals (meaning the intersection of individual level surfaces) is just the (unparameterised) integral curve.  Well known elementary examples are the isotropic harmonic oscillator, the Kepler system and the Calogero-Moser system.

The quadratures of complete integrability are often achieved through the separation of variables of the Hamilton-Jacobi equation.  The solution of a maximally super-integrable system can also be calculated purely algebraically (albeit implicitly), requiring just the solution of the equations $I_k=c_k,\, k=1,\dots ,2n-1$.  Maximally superintegrable systems have a number of interesting properties: they can be separable in more than one coordinate system; all bounded orbits are closed; they give rise to interesting Poisson algebras with polynomial Poisson relations.

The idea can be extended to {\em quantum integrable systems}, with first integrals replaced by commuting differential operators.  If we are truly interested in the system as a physical quantum system, then we would require our operators to be Hermitian.  However, the same ideas are applicable to the study of polynomial eigenfunctions of Laplace-Beltrami operators, so we will not impose this condition.  In fact, algebraically, Poisson algebras and operator algebras are {\em formally} very similar and it is very straightforward to ``quantise'' many classical, super-integrable systems.  For some examples of superintegrable quantum systems it is possible to use the additional commuting operators to {\em explicitly} build sequences of eigenfunctions \cite{f05-1,f06-1,f07-1}, as will be seen below.

There is a large literature on the classification and analysis of superintegrable systems (see the review \cite{13-2}) and they naturally occur in many applications in physics (additional integrals being referred to as ``hidden symmetries'' \cite{14-2}).

Section \ref{ResHarmonic} is expository, using the well known example of resonant harmonic oscillators to explain the basics of super-integrability.  It will be seen how the classical and quantum versions are more or less just two different representations of a single algebraic structure.

In Section \ref{concurv} we consider Hamiltonians in ``natural form'' (the sum of kinetic and potential energies).  The kinetic energy has geometric origins, and, in the flat and constant curvature cases, the existence of a large isometry group leads to the superintegrability of the geodesic equations.  We emphasise the role of these isometries in the construction of separable potentials and in the choice of additional integrals for the superintegrable case.  These ideas are illustrated through a 3D constant curvature case, introduced in \cite{f17-5}, with the symmetry algebra and its {\em involutive automorphisms} playing an important role in constructing a $10-$dimensional Poisson algebra.

Section \ref{sec:zernike} is concerned with the classical and quantum Zernike system, whose super-integrability (and much more!) was recently studied in detail by Pogosyan, et al \cite{17-1,17-2}.  My purpose here is show how the methods of Section \ref{concurv} can be applied, giving a new perspective on their results.  The quantum case is equivalent to case IX of the Krall-Sheffer classification of 2D second order, linear differential operators with polynomial eigenfunctions \cite{67-2,f13-1}.

\section{Resonant Harmonic Oscillators}\label{ResHarmonic}

This section is expository, using a well known example to explain the basics of super-integrability.  For this simple example, all information can be gleaned from the explicit solution, which can be found by elementary means.  However, we can extract the same information from additional first integrals, which exist in the resonant case.  This sledge hammer can then be applied to much tougher nuts, whose explicit solution is far from elementary.

As usual, the object of the classical game is the trajectory, whilst that of the quantum case is the spectrum and the eigenfunctions.  However, most of the intermediate steps are {\em algebraically} identical, with one in terms of Poisson algebras and the other in operator algebras.  In the classical case, the first integrals are used to construct an algebraic curve, which represents the familiar Lissajous figure.  In the quantum case, the existence of commuting operators accounts for the degeneracy of an eigenvalue.  The commuting operators act like ``horizontal ladders'' taking us through the eigenspace of a given eigenvalue.

\subsection{The Classical Case}

Consider the system with Hamiltonian
$$
H=H_1+H_2, \quad\mbox{where}\quad  H_k=\frac{1}{2} (p_k^2+\omega_k^2 q_k^2), \quad\mbox{with}\quad  \{H_1,H\}=\{H_2,H\}=0.
$$
As shown by Jauch and Hill \cite{40-2}, the classical (not just the quantum) oscillator has ladder operators.  By writing
$$
p_k^2+\omega_k^2 q_k^2 = (p_k+i\omega_k q_i)(p_k-i\omega_k q_k),
$$
we find
\be\label{ladder-class}
\{p_k+i\omega_k q_k,p_k-i\omega_k q_k\}=2i\omega_k\quad\Rightarrow\quad    \{H_k,p_k\mp i \omega_k q_k\}=\pm i \omega_k (p_k\mp i \omega_k q_k).
\ee
Generically the ratio $\omega_1/\omega_2$ is irrational, so the trajectory performs {\em quasi-periodic motion}, filling the torus, defined by $H_i=\frac{1}{2} \omega_i^2 A_i^2$, where $A_i$ are the initial amplitudes.

In the resonant case, with $\omega_1=m,\, \omega_2=n$ (integers), the solution is periodic and, projecting onto the $(q_1,q_2)$ plane, the trajectory forms a Lissajous figure, which represents an algebraic curve in the plane.

In the resonant case, (\ref{ladder-class}) implies
$$
\{H,(p_1\mp i m q_1)^n\} = \pm imn (p_1\mp i m q_1)^n \quad\mbox{and}\quad  \{H,(p_2\mp i n q_2)^m\} = \pm imn (p_2\mp i n q_2)^m,
$$
so
$$
K=(p_1+miq_1)^n(p_2-niq_2)^m=K_1+iK_2
$$
gives us a pair of {\em real} first integrals.  For low values of $m$ and $n$ these are easy to write explicitly, but are generally very complicated as functions of $q_k, p_k$.  However, the general structure in terms of ladder operators enables us to deduce the complete set of Poisson relations for $H_1, H_2, K_1, K_2$.  We already have $\{H_1,H_2\}=0$ and $H=H_1+H_2$, but then
\bea
&&  \{H_1,K_1\}=-\{H_2,K_1\} = m n K_2,\quad   \{H_1,K_2\}=-\{H_2,K_2\} = -m n K_1,  \nn\\
&&   \{K_1,K_2\} = 2 m n (2 H_1)^{n-1} (2 H_2)^{m-1} (m H_1-n H_2) \quad\mbox{and}\quad K_1^2+K_2^2 = (2 H_1)^n (2 H_2)^m,  \nn
\eea
the latter showing that at most $3$ of these integrals are independent.  It is easy to see that $H_1, H_2$ and $K_1$ are functionally independent.

\paragraph{Lissajous Figures:}  The general solution of the equations of motion is given by $q_1=A_1\sin(m t+\delta_1),\;\; q_2=A_2\sin(n t+\delta_2)$, with the constants $(A_1,A_2,\delta_1,\delta_2)$ being determined by initial conditions.  A parametric plot then gives the Lissajous figure.

The values of the four integrals $H_i(0)=h_i,\, K_i(0)=k_i$, are fixed by the initial conditions and satisfy $k_1^2+k_2^2=(2 h_1)^n(2 h_2)^m$.  We can then use three of these integrals to eliminate $p_i$ and obtain an algebraic relation between $q_1$ and $q_2$.  This gives a {\em non-parametric} representation of the Lissajous curve.

\bex[The Case $(m,n)=(1,2)$]\label{ex-res}  {\em
Here we have $H_1=\frac{1}{2} (p_1^2+q_1^2),\, H_2=\frac{1}{2} (p_2^2+4q_2^2)$ and
\be\label{k1k2-res-harm}
K_1 = p_1^2p_2-q_1^2p_2+4q_1q_2p_1, \quad K_2 = 2 q_1p_1p_2-2 q_2p_1^2+2 q_1^2q_2.
\ee
The solution $q_1= \sin t, q_2 = 2 \sin (2 t), p_1 = \cos t, p_2 = 4 \cos (2 t)$, with $q_1(0)=q_2(0)=0, p_1(0)=1, p_2(0)=4$, corresponds to Figure \ref{w1=1w2=2}.  The first integrals have values $H_1=\frac{1}{2},\, H_2=8,\, K_1=4,\, K_2=0$, so
$$
p_1^2=1-q_1^2,\; p_2^2 = 4 (4-q_2^2),\; q_1p_1p_2 = q_2(p_1^2-q_1^2)= q_2 (1-2 q_1^2)\quad\Rightarrow\quad q_2^2 = 16 q_1^2 (1-q_1^2),
$$
giving the algebraic curve corresponding to Figure \ref{w1=1w2=2}.
\begin{figure}[htb]
\centering
 \subfigure[Symmetric Figure of Eight]{
\includegraphics[width=5cm,height=3cm]{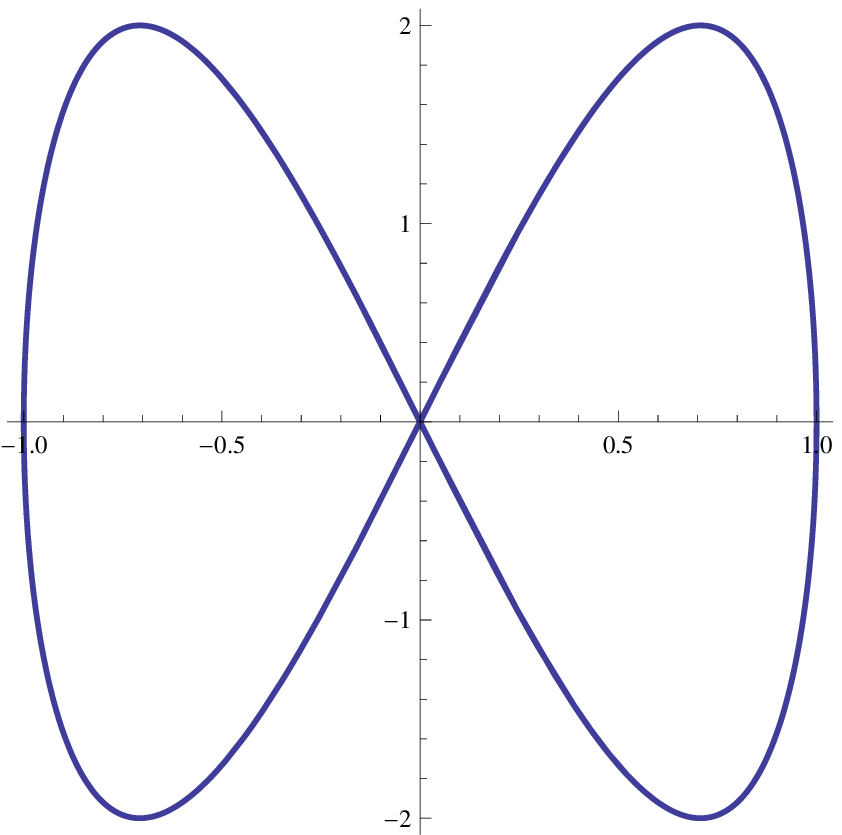}\label{w1=1w2=2}
} \qquad \subfigure[Degenerate Parabolic Figure]{\includegraphics[width=5cm,height=3cm]{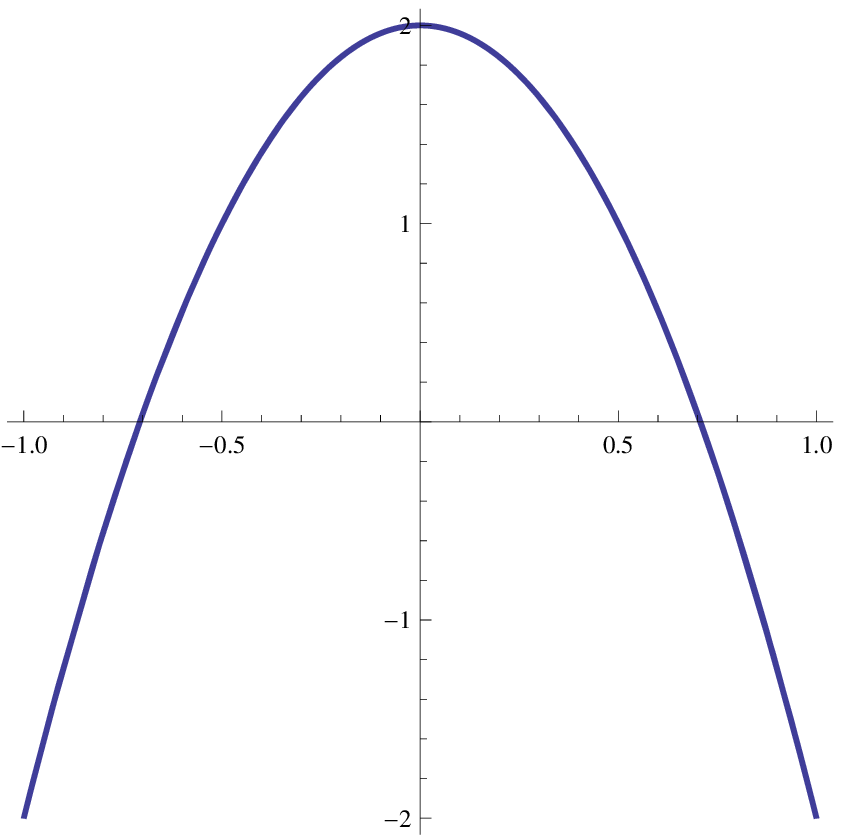}\label{w1=1w2=2-degen}
}\caption{Lissajous Figures with $(m,n)=(1,2)$}\label{Lissajous}
\end{figure}

The solution $q_1= \sin t, q_2 = 2 \cos (2 t), p_1 = \cos t, p_2 = -4 \sin (2 t)$, with $q_1(0)=0, q_2(0)=2, p_1(0)=1, p_2(0)=0$, corresponds to Figure \ref{w1=1w2=2-degen}.  The first integrals have values $H_1=\frac{1}{2},\, H_2=8,\, K_1=0,\, K_2=-4$, which leads to the degenerate form: $q_2^2=4 (1-2 q_1^2)^2$.  Clearly, Figure \ref{w1=1w2=2-degen} corresponds to $q_2=2 (1-2 q_1^2)$.
}\eex

\subsection{The Quantum Case}\label{QHO}

The quantum version of the harmonic oscillator is algebraically very similar.  We consider the eigenvalue problem
$$
L\psi\equiv (\pa_x^2+\pa_y^2-\omega_1^2 x^2-\omega_2^2 y^2) \psi = \lambda \psi,
$$
with ladder operators $A_x^\pm = \mp \pa_x + \omega_1 x,\, A_y^\pm = \mp \pa_y + \omega_2 y$, satisfying
$$  
[L,A_x^\pm]=\mp 2 \omega_1 A_x^\pm,\quad [L,A_y^\pm]=\mp 2 \omega_2 A_y^\pm,\quad [A_x^\pm,A_y^\pm]=0.
$$   
The ground state $\psi_{00}=e^{-\frac{1}{2}(\omega_1x^2+\omega_2y^2)}$ satisfies
$$
A_x^-\psi_{00}=A_y^-\psi_{00}=0,\quad L\psi_{00}= -(\omega_1+\omega_2)\psi_{00},
$$
and we use the raising operators to define an infinite triangular array of eigenfunctions:
$$
\psi_{ij} = (A_x^+)^i (A_y^+)^j \psi_{00}, \quad\mbox{with eigenvalues}\;\; \lambda_{ij}=-(2i+1)\omega_1-(2j+1)\omega_2.
$$
\begin{figure}[hbt]
\begin{center}
\unitlength=0.5mm
\begin{picture}(160,80)
\put(80,80){\makebox(0,0){$(0,0)$}}
\put(60,60){\makebox(0,0){$(1,0)$}}
\put(100,60){\makebox(0,0){$(0,1)$}}
\put(40,40){\makebox(0,0){$(2,0)$}}
\put(80,40){\makebox(0,0){{$(1,1)$}}}
\put(120,40){\makebox(0,0){$(0,2)$}}
\put(20,20){\makebox(0,0){$(3,0)$}}
\put(60,20){\makebox(0,0){$(2,1)$}}
\put(100,20){\makebox(0,0){$(1,2)$}}
\put(140,20){\makebox(0,0){$(0,3)$}}
\put(0,0){\makebox(0,0){$(4,0)$}}
\put(40,0){\makebox(0,0){{$(3,1)$}}}
\put(80,0){\makebox(0,0){$(2,2)$}}
\put(120,0){\makebox(0,0){{$(1,3)$}}}
\put(160,0){\makebox(0,0){$(0,4)$}}
\put(75,75){\vector(-1,-1){10}}
\put(85,75){\vector(1,-1){10}}\put(65,55){\vector(1,-1){10}}\put(95,55){\vector(-1,-1){10}}
\put(55,55){\vector(-1,-1){10}}\put(45,35){\vector(1,-1){10}}  \put(105,55){\vector(1,-1){10}}
\put(35,35){\vector(-1,-1){10}} \put(75,35){\vector(-1,-1){10}} \put(115,35){\vector(-1,-1){10}} \put(125,35){\vector(1,-1){10}}\put(25,15){\vector(1,-1){10}}\put(85,35){\vector(1,-1){10}}\put(65,15){\vector(1,-1){10}}\put(105,15){\vector(1,-1){10}}
\put(15,15){\vector(-1,-1){10}}\put(55,15){\vector(-1,-1){10}} \put(95,15){\vector(-1,-1){10}} \put(135,15){\vector(-1,-1){10}} \put(145,15){\vector(1,-1){10}}
\put(40,60){\makebox(0,0){$A_x^+$}}
\put(120,60){\makebox(0,0){$A_y^+$}}
\end{picture}
\end{center}
\caption{The action of $A_x^+$ and $A_y^+$ on the array $\psi_{(jk)}$}\label{2dHarray}
\end{figure}
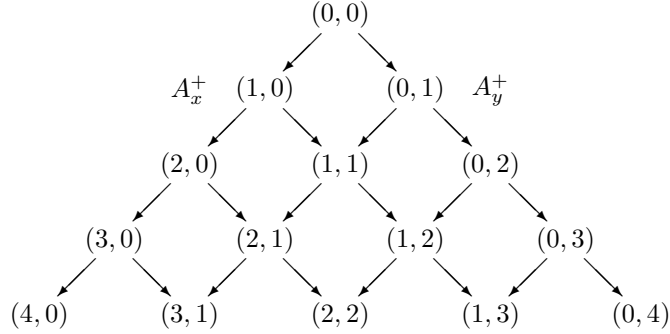
Since the ladders in the $x-$direction commute with those in the $y-$direction, it doesn't matter which order we operate with $A_x^+$ and $A_y^+$.  Each order corresponds to a different path through the array depicted in Figure \ref{2dHarray}.  We can also move in the negative direction $A_x^-\psi_{jk}=2 i \omega_1 \psi_{j-1\,k},\, A_y^-\psi_{jk}=2 i \omega_2 \psi_{j\,k-1}$.

Furthermore, we have $\psi_{jk} = \psi_{00} H_j(x)H_k(y)$, where $H_k$ are just the Hermite polynomials (but with coefficients depending upon $\omega_1$ or $\omega_2$).  Once again everything we need is obtained by elementary calculations.

\medskip
{\em The eigenvalues $\lambda_{ij}$ are distinct, if and only if $\omega_1$ and $\omega_2$ are {\em not} commensurate.}

\medskip
When $(\omega_1,\omega_2)=(m,n)$, we can build two commuting operators:
$$
[L,(A_x^\pm)^n] = \mp 2mn (A_x^\pm)^n,\quad [L,(A_y^\pm)^m] = \mp 2mn (A_y^\pm)^m\quad\Rightarrow\quad \left\{\begin{array}{l}
                                                                                                               \left[L, (A_y^+)^m (A_x^-)^n \right] = 0 ,\\[2mm]
                                                                                                                \left[L,(A_x^+)^n (A_y^-)^m\right] = 0.
                                                                                                              \end{array}\right.
$$
These play the role of connecting eigenfunctions with the {\em same} eigenvalue, so are directly related to {\em degeneracy}.

\bex[The Isotropic Oscillator]  {\em
When $\omega_1=\omega_2=1$, we have $\lambda_{ij}=-2(i+j+1)$, so, for each integer $\ell\geq 0$, we have $\ell+1$ pairs $(i,j)$ satisfying $i+j=\ell$, corresponding to eigenfunctions with eigenvalue $\lambda_{ij}=-2(\ell+1)$.  These lie on the horizontal levels in Figure \ref{2dHarray}.  The operators
$$
A_y^+A_x^- = x y+y\pa_x-x\pa_y-\pa_x\pa_y    \quad\mbox{and}\quad     A_x^+A_y^- = x y-y\pa_x+x\pa_y-\pa_x\pa_y
$$
move us, respectively, right and left across the horizontal levels.  The usual commuting operators for this case are just
$$
K_1=\frac{1}{2}(A_x^-A_y^+ - A_x^+A_y^-)=y\pa_x-x\pa_y,\quad  K_2=-\frac{1}{2}(A_x^-A_y^+ + A_x^+A_y^-)=\pa_x\pa_y-x y.
$$
}\eex

\bex[The Case $(m,n)=(1,2)$]  {\em
In this case, $\lambda_{ij}=-(2(i+2j)+3)$, so, starting on the left edge, with $(i,j)=(m,0)$, the eigenfunctions with eigenvalue $\lambda_{m\, 0}=-(2m+3)$, satisfy $i+2j=m$.  For $m=2\ell$ or $m=2\ell+1$, we have $\ell+1$ solutions.  The operators $(A_y^+ A_x^-)^2$ and $(A_x^+)^2A_y^-$ move us respectively right and left in the direction of the line $i+2j=m$.  For example, the first of these takes the point $(i,j)$ to $(i-2,j+1)$ (or, more precisely, $\psi_{ij}\mapsto\psi_{i-2\, j+1}$).

Operators which look more like the integrals of (\ref{k1k2-res-harm}) are the even and odd parts
\bea
&& K_1=\frac{1}{2} ((A_x^+)^2A_y^- + (A_x^-)^2A_y^+)= 2y\pa_x^2 -2x \pa_x\pa_y-\pa_y+2x^2y,\nn\\
&&   K_2=\frac{1}{2} ((A_x^+)^2A_y^- - (A_x^-)^2A_y^+)= \pa_x^2\pa_y -4x y  \pa_x+x^2\pa_y-2y,\nn
\eea
which satisfy
\bea
&&   [K_1,L_1]=4 K_2,\quad [K_1,L_2]=-4 K_2,\quad [K_2,L_1]=4 K_1,\quad [K_2,L_2]=-4 K_1,\nn\\
&&  [K_1,K_2]= 4 L_1 L_2 -2 L_1^2 -6,\quad  K_1^2-K_2^2+L_1^2L_2=8L_1-3L_2.  \nn
\eea
}\eex

\br[Separability]
For any values of $\omega_i$, this system is clearly separable in Cartesian coordinates.  However, as is a feature of super-integrability, each of the above resonant cases is separable in a second coordinate system.  The isotropic case is separable in polar coordinates, whilst the case $(m,n)=(1,2)$ is separable in parabolic coordinates.
\er

\section{Motion on Flat and Constant Curvature Manifolds}\label{concurv}

In this section we consider Hamiltonians in ``natural form'' (the sum of kinetic and potential energies).  A non-degenerate kinetic energy is associated with (pseudo-)Riemannian metric, so we first consider the geodesic equations on such manifolds.  We are particularly interested in the flat and constant curvature manifolds, in which case the geodesic motion is completely integrable, and even maximally super-integrable.  We then consider the problem of adding potentials in such a way that the Hamiltonian system is still completely integrable.  Since we restrict ourselves to integrals which are {\em quadratic} in momenta, we naturally build {\em separable systems}, depending on arbitrary functions.  We then ask if it is possible to restrict these arbitrary functions in order to build {\em super-integrable} systems.

\subsection{The Geodesic Equations}

For a manifold with coordinates $(q_1,\dots ,q_n)$, metric coefficients $g_{ij}$, with inverse $g^{ij}$, the geodesic equations are Hamiltonian, with {\em kinetic energy}
\be\label{Ham-h2}
H = \frac{1}{2}\, \sum_{i,j=1}^n g^{ij}p_ip_j,\quad\mbox{where}\quad  p_i=\sum_k g_{ik}\dot q_k.
\ee
For a metric with isometries, the infinitesimal generators (Killing vectors) give rise to first integrals, which are {\em linear} in momenta (Noether constants).
When the space is either flat or constant curvature, it possesses the maximal group
of isometries, which is of dimension $\frac{1}{2}n(n+1)$.  In this case, (\ref{Ham-h2}) is actually the second order {\em Casimir} function of the symmetry algebra (see \cite{74-7}).

\br[Maximally Super-Integrable]
Notice that since $\frac{1}{2}n(n+1)-(2n+1) = \frac{1}{2}(n-1)(n-2)$, the geodesic equations on such spaces are always maximally super-integrable and that, when $n\geq 3$, the Noether constants cannot be {\em functionally independent} (the Killing vectors are, of course, {\em linearly} independent). The standard example is the Euclidean algebra in 3D, for which ${\bf p}\cdot {\bf L}=0$.
\er

\subsubsection{An Explicit Constant Curvature Metric}

As a working example, we now consider the Hamiltonian
\be\label{H12}
H =  q_1^2(p_1^2-p_2^2-p_3^2),
\ee
corresponding to the diagonal upper index metric ${\rm{diag}}\left(q_1^2,-q_1^2,-q_1^2 \right)$, which has constant curvature, with isometry algebra (see \cite{f17-5})
\bea
&& e_1 = p_2, \quad h_1 = -2(q_1p_1+q_2p_2+q_3p_3), \quad  f_1 = -2q_1q_2p_1+(q_3^2-q_1^2-q_2^2)p_2-2q_2q_3p_3 ,\nn\\[-2mm]
&&  \label{diag-alg}\\[-2mm]
&& e_2 = p_3, \quad h_2 = 2(q_2p_3-q_3p_2), \quad  f_2 = -4q_3(q_1p_1+q_2p_2)-2(q_1^2-q_2^2+q_3^2)p_3 , \nn
\eea
with Poisson relations given in Table \ref{Tab:6Symm_alg}.  We see that it has rank 2.  In fact, it is easy to see that this is isomorphic to $\mathbf{so}(1,3)$.
\begin{table}[h]\centering
\caption{The $6-$dimensional isometry algebra of (\ref{H12})}\label{Tab:6Symm_alg}\vspace{3mm}
\begin{tabular}{|c||c|c|c||c|c|c|}
\hline   &$e_1$    &$h_1$     &$f_1$   &$e_2$   &$h_2$    &$f_2$    \\[.10cm]\hline\hline
$e_1$    &0        &$2e_1$    &$-h_1$  &0       &$-2e_2$    &$-2h_2$      \\[1mm]\hline
$h_1$    &$-2e_1$   &0        &$2f_1$  &$-2e_2$ &0        &$2f_2$   \\[1mm]\hline
$f_1$    &$h_1$    &$-2f_1$   &0       &$-h_2$  &$-f_2$    &0        \\[1mm]\hline\hline
$e_2$    &0         &$2e_2$   &$h_2$ &0       & $2e_1$   & $-2h_1$   \\[1mm]\hline
$h_2$    &$2e_2$    &0        &$f_2$  &$-2 e_1$   &0        &$-4 f_1$    \\[1mm]\hline
$f_2$    &$2h_2$   &$-2f_2$  &0       &$2 h_1$   &$4 f_1$   &0        \\[1mm]\hline
\end{tabular}
\end{table}

The quadratic Casimir of this algebra is proportional to $H$:
$$
H = e_1f_1+\frac{1}{4} h_1^2+ \frac{1}{4} (2e_2 f_2-h_2^2).
$$
There is a second independent (fourth order) Casimir element of the abstract algebra, but in this representation it is a perfect square and {\em zero}, giving the quadratic constraint
$e_1 f_2+h_1 h_2-2 f_1 e_2=0$.
This algebra has the following useful pair of involutive automorphisms:
\bea
 \iota_1:\;  (e_1,h_1,f_1,e_2,h_2,f_2)  &\mapsto &   \left(f_1,-h_1,e_1,-\frac{1}{2} f_2,-h_2,-2 e_2\right),  \nn\\
 \iota_{23}:\; (e_1,h_1,f_1,e_2,h_2,f_2) &\mapsto & \left(e_2,h_1,\frac{1}{2} f_2,e_1,-h_2,2 f_1\right).  \nn
\eea
The involution $\iota_{23}$ just corresponds to the interchange $(q_2,p_2) \leftrightarrow (q_3,p_3)$, so is clearly a canonical transformation.  The involution $\iota_1$ is also canonical, with generating function $S_1 = \frac{q_1P_1-q_2P_2+q_3P_3}{q_1^2-q_2^2-q_3^2}$.  {\em These will be important later}.

\subsection{Adding Potentials: Separability}\label{separable}

We start with a kinetic energy $H_0$ corresponding to a flat or constant curvature metric and use its isometry algebra to build commuting quadratic (in momenta) functions.  We can then add potential functions.  This will be explained in the context of the specific 3D Hamiltonian (\ref{H12}), but the idea is easily applied to any metric with enough symmetries.

Starting with $H = H_0 + h({\bf q})$, with kinetic energy $H_0$ of (\ref{H12}), we seek two functions $F_1$ and $F_2$, such that $H, F_1, F_2$ are in involution:
$$   
\{H,F_1\} = \{H,F_2\} = \{F_1,F_2\} = 0.
$$   
In our quadratic case, such functions will be the sum of two homogeneous parts, $F_i=F^{(2)}_i + F^{(0)}_i$, and
$$
\{H,F_i\}=0 \quad\Rightarrow\quad \{H_0,F^{(2)}_i\}=0 \quad\mbox{and}\quad \{H_0,F^{(0)}_i\}+\{h,F^{(2)}_i\} = 0.
$$
The first of these means that the coefficients of $p_i p_j$ in $F^{(2)}_i$ define a second order Killing tensor of the metric corresponding to $H_0$.  When this metric is constant curvature, \underline{all} Killing tensors are built as tensor products of Killing vectors (see \cite{74-7}).  In the Poisson representation, this just means that $F^{(2)}_i$ is some quadratic form of the elements of the isometry algebra.  Since this algebra is of rank $2$, any isometry $K$ will commute with exactly one other element $\bar K$.  Since we require $\{F^{(2)}_1,F^{(2)}_2\}=0$, we choose these quadratic parts to be independent quadratic forms of some pair $K, \bar K$.  For simplicity, we can choose our pairs to be one of $e_1,e_2$, or $h_1,h_2$ or $f_1,f_2$.

The choice of quadratic integrals means that our systems will be separable.  The calculation of separable potentials is standard (in principle!) and it is well known that in the standard orthogonal coordinate systems, with separable kinetic energies, we can add potentials which depend upon a number of arbitrary functions of a single variable \cite{76-8}.  If a complete (possessing $n$ parameters) solution of the Hamilton-Jacobi equation is found, then, by Jacobi's theorem, these parameters, when written in terms of the canonical variables, are quadratic (in momenta) first integrals of $H$.  The problem has also been posed in the ``opposite'' direction: given a pair of Poisson commuting, homogeneously quadratic integrals (in two degrees of freedom) what sort of potentials can be added, whilst maintaining commutativity?  This is a classical problem (see Whittaker \cite{88-4}, chapter 12, section 152) and leads to the Bertrand-Darboux equation for the potential \cite{88-8,08-7}.

\subsubsection*{The Commuting Pairs $e_1, e_2$ and $f_1, f_2$}

These two cases are connected by the action of the automorphism $\iota_1$.  It should be emphasised that without the knowledge of the underlying isometry algebra, these two cases would be {\em independent}, with the $f_1,f_2$ case presenting a considerably more difficult calculation!

Starting with the pair $e_1, e_2$, we consider the two quadratic integrals
\begin{subequations}\label{ee}
\be\label{e^2}
F_1 = e_1^2+g_1({\bf q})= p_2^2+g_1({\bf q}),\quad F_2 = e_2^2+g_2({\bf q})= p_3^2+g_2({\bf q}).
\ee
The simple form of $e_1$ and $e_2$ means that we are already in separation coordinates, leading to
\be\label{eesols1}
h = -q_1^2(\varphi_1(q_2)+\varphi_2(q_3))+\varphi_3(q_1),\quad g_1 = \varphi_1(q_2),\quad g_2 = \varphi_2(q_3).
\ee
\end{subequations}

The much more difficult case to calculate, involving $f_1$ and $f_2$, is simply obtained by using the automorphism $\iota_1$, which preserves $H_0$, whilst mapping $e_1^2\mapsto f_1^2$ and $e_2^2\mapsto \frac{1}{4} f_2^2$.  Recall that this involution is realised by the {\em canonical transformation}, generated by
$$
S_1 = \frac{q_1P_1-q_2P_2+q_3P_3}{q_1^2-q_2^2-q_3^2} \quad\Rightarrow\quad Q_1=\frac{q_1}{q_1^2-q_2^2-q_3^2},\;\; Q_2=\frac{-q_2}{q_1^2-q_2^2-q_3^2},\;\; Q_3=\frac{q_3}{q_1^2-q_2^2-q_3^2}.
$$
This gives
\begin{subequations}
\bea
&& F_1 = f_1^2+g_3({\bf q})= (-2q_1q_2p_1+(q_3^2-q_1^2-q_2^2)p_2-2q_2q_3p_3)^2+g_3({\bf q}),\nn\\[-2mm]
                             \label{f^2}         \\[-2mm]
&& F_2 = \frac{1}{4}f_2^2+g_4({\bf q})= (2q_3(q_1p_1+q_2p_2)+(q_1^2-q_2^2+q_3^2)p_3)^2+g_4({\bf q}),\nn
\eea
where
\bea
&& g_3 = \varphi_1\left(\frac{-q_2}{q_1^2-q_2^2-q_3^2}\right), \quad g_4 = \varphi_2\left(\frac{q_3}{q_1^2-q_2^2-q_3^2}\right), \nn\\[-2mm]
                             \label{g34}         \\[-2mm]
&&  h = \varphi_3\left(\frac{q_1}{q_1^2-q_2^2-q_3^2}\right) - \frac{q_1^2 \left(\varphi_1\left(\frac{-q_2}{q_1^2-q_2^2-q_3^2}\right)+\varphi_2\left(\frac{q_3}{q_1^2-q_2^2-q_3^2}\right)\right)}{(q_1^2-q_2^2-q_3^2)^2}.\nn
\eea
\end{subequations}

\subsection{Adding More Potentials: Super-integrability}\label{super}

We now suppose we have an involutive system $H, F_1, F_2$, as above.  For super-integrability, we must add two more functions $F_3,F_4$, satisfying $\{H,F_3\}=\{H,F_4\}=0$, but with the (now) {\em given} $H$.  The functions should be chosen to be {\em functionally independent}, so the Jacobian matrix
$$
\frac{\pa(H,F_i)}{\pa {\bf x}},\quad\mbox{where}\quad   {\bf x} = (q_1,\dots , p_3),
$$
has \underline{maximal rank} $5$, since in this case, the level surface ${\cal S} = \{{\bf x}: H=c_0,F_i=c_i\}_{i=1}^4$, has dimension \underline{one}, so represents an (unparameterised) trajectory of the dynamical system.  In the case of the resonant oscillator of Example \ref{ex-res}, this enabled us to derive the algebraic curve for a Lissajous figure, but generally, we cannot reduce to such explicit formulae.

\subsubsection*{The case with $F_1=e_1^2+g_1,F_2=e_2^2+g_2,F_3=f_1^2+g_3,F_4=\frac{1}{4}f_2^2+g_4$}\label{sec:eifi}

Here we start with the formulae of (\ref{ee}) and seek a pair of additional functions $F_3 = f_1^2+g_3({\bf q})$ and $F_4 = \frac{1}{4}f_2^2+g_4({\bf q})$, satisfying $\{H,F_3\}=\{H,F_4\}=0$.  This restricts our potentials from depending upon 2 {\em arbitrary functions} to 2 {\em arbitrary parameters}:
$$  
h=q_1^2\, \left(\frac{k_1}{q_2^2}+\frac{k_2}{q_3^2}\right), \;\; g_1 = -\,\frac{k_1}{q_2^2},  \;\; g_2 = -\,\frac{k_2}{q_3^2}, \;\;
                    g_3 = -\,\frac{k_1 (q_2^2+q_3^2-q_1^2)^2}{q_2^2},\;\;  g_4 = -\,\frac{k_2 (q_2^2+q_3^2-q_1^2)^2}{q_3^2}.
$$  
In this case, we also find that $h_1$ is a first integral.  We therefore have $6$ first integrals $(H,F_1,F_2,F_3,F_4,h_1)$ (with Jacobian of rank $5$) and find
$$
\{F_1,F_2\}=\{F_3,F_4\}=0\;\;\;\mbox{and}\;\;\; \{F_i,h_1\}=\lambda_i F_i,\;\; i=1,\dots ,4,\;\;\mbox{where}\;\; \lambda = (4,4,-4,-4).
$$
Under the action of the involutions $\iota_1$ and $\iota_{23}$, we have
\bea
\small \iota_1:(H,F_1,F_2,F_3,F_4,h_1,k_1,k_2) &\mapsto & (H,F_3,F_4,F_1,F_2,-h_1,k_1,k_2),\nn\\
\iota_{23}:(H,F_1,F_2,F_3,F_4,h_1,k_1,k_2) &\mapsto & (H,F_2,F_1,F_4,F_3,h_1,k_2,k_1).\nn
\eea
The Poisson brackets $\{F_1,F_3\},\, \{F_1,F_4\},\, \{F_2,F_3\},\, \{F_2,F_4\}$ are all cubic in momenta, but cannot be written as linear combinations of $\{h_1 F_i,h_1 H, h_1^3\}_{i=1}^4$, so we need to introduce new elements.
Since the involutions are canonical, $\{F_1,F_3\}$ and $\{F_2,F_4\}$ (respectively $\{F_1,F_4\}$ and $\{F_2,F_3\}$) are related through the involutions.  Since $\{F_1,F_3\}$ and $\{F_2,F_4\}$ factorise, we can define two new \underline{quadratic} elements $F_5, F_6$ through the relations
$$
\{F_1,F_3\} = h_1(h_1^2-4H-4F_5-4k_1),\quad  \{F_2,F_4\} = h_1(h_1^2-4H-4F_6-4k_2),
$$
with $F_5\leftrightarrow F_6$ under $\iota_{23}$.  We define $F_7,F_9$ by the equations
$$
\{F_1,F_6\}= 2 h_1 F_1+4 F_7,\quad \{F_3,F_6\}= -2 h_1 F_3+4 F_9,
$$
related by $F_7\leftrightarrow F_9$ under $\iota_{1}$.

The function $F_8$ is defined by the second of the following equations
$$
\{F_1,F_4\}+\{F_2,F_3\}= 8h_1\left(H+F_5+F_6-\frac{1}{4}\, h_1^2\right),\quad   \{F_1,F_4\}-\{F_2,F_3\}= 16 F_8,
$$
after which, we find that $\{F_5,F_6\}=4 F_8$.

\begin{figure}[hbt]
\unitlength=0.65mm
\centering
\subfigure[Four connected bracket relations]{\begin{picture}(100,50)\thicklines
\put(0,25){\makebox(0,0){$P_{16}$}} \put(50,0){\makebox(0,0){$P_{25}$}}
\put(50,50){\makebox(0,0){$P_{36}$}} \put(100,25){\makebox(0,0){$P_{45}$}}
\put(5,25){\vector(2,1){40}} \put(5,25){\vector(2,-1){40}}
\put(55,5){\vector(2,1){35}} \put(55,45){\vector(2,-1){35}}
\put(25,40){\makebox(0,0){$\iota_1$}} \put(25,10){\makebox(0,0){$\iota_{23}$}}
\put(75,40){\makebox(0,0){$\iota_{23}$}} \put(75,10){\makebox(0,0){$\iota_1$}}
\end{picture}\label{F1636}}
\qquad \subfigure[Two connected bracket relations]{\begin{picture}(100,50)\thicklines
\put(25,25){\makebox(0,0){$P_{13}$}} \put(75,25){\makebox(0,0){$P_{24}$}}
\put(30,25){\vector(2,0){40}}
 \put(50,30){\makebox(0,0){$\iota_{23}$}}
\end{picture}\label{F1324}}
\caption{Bracket relations connected through $\iota_1$ and $\iota_{23}$} \label{F16361324}
\end{figure}
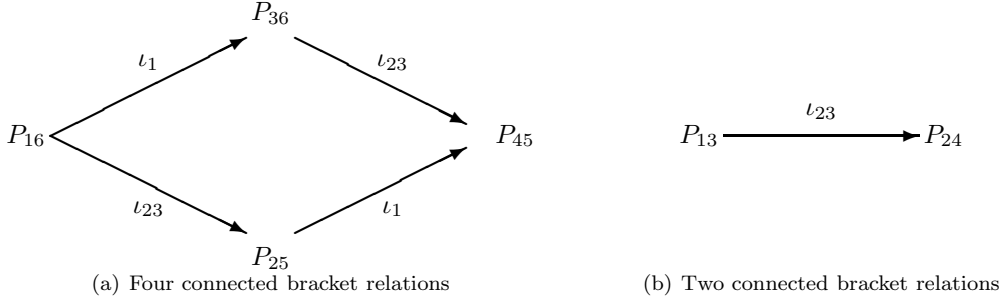

In this way we build a $10-$dimensional Poisson algebra.  The action of the two involutions is then given by:
$$
\begin{array}{|c||c|c|c|c|c|c|c|c|c|c|c|c|c|}\hline
& H & F_1 & F_2 & F_3 & F_4 & F_5 & F_6 & F_7 & F_8 & F_9 & h_1 & k_1 & k_2 \\\hline
\iota_1: & H & F_3 & F_4  & F_1 & F_2 & F_5 & F_6 & F_9 & F_8 & F_7 & -h_1 & k_1 & k_2\\\hline
\iota_{23}: & H & F_2 & F_1  & F_4 & F_3 & F_6 & F_5 & -F_7 & -F_8 & -F_9 & h_1 & k_2 & k_1\\ \hline
\end{array}
$$
The action of $\iota_{23}$ on $\{F_1,F_6\}$ and $\{F_3,F_6\}$ then gives
$$
\{F_2,F_5\}= 2 h_1 F_2-4 F_7,\quad \{F_4,F_5\}= -2 h_1 F_4-4 F_9.
$$
This phenomenon of connecting four different Poisson relations through the involutions is depicted in Figure \ref{F1636}, where we define $P_{ij}=\{F_i,F_j\}$.
Sometimes only \underline{two} relations are connected, such as with $P_{13}$ and $P_{24}$, or even just \underline{one}, such as with $P_{79}$, because of invariance properties.  These relations mean that when we find one entry $P_{ij}$, we automatically have up to 3 others.

\br[Nontrivial Relations]
Out of our $11$ functions, only $5$ can be functionally independent.  The Jacobi identity, applied to our $10$ functions $F_i$ and $h_1$, give us the necessary, nontrivial relations.
\er

\subsubsection*{The Trajectory as a Level Curve}

Consider the level curve, defined by $h_1=c_0$ and $\{F_i=c_i\}_{i=1}^4$.  By omitting either $F_4$ or $F_3$, we can eliminate $p_i$ in two ways, to get
$$
(c_1\eta^2-c_0^2q_2^2)^2-2(2 k_1c_0^2+c_1c_3) \eta^2+c_3(c_3-2 c_0^2 q_2^2)=0,\quad   (c_2\eta^2-c_0^2q_3^2)^2-2(2 k_2c_0^2+c_2c_4) \eta^2+c_4(c_4-2 c_0^2 q_3^2)=0,
$$
where $\eta=q_1^2-q_2^2-q_3^2$.  These are related through the action of $\iota_{23}$.  The special case of $c_3=c_4=0$ gives
\begin{subequations}
\be\label{eta2}
c_1\eta^2-c_0^2q_2^2 = \pm 2 \sqrt{k_1}c_0\eta,\qquad   c_2\eta^2-c_0^2q_3^2 = \pm 2 \sqrt{k_2}c_0\eta,
\ee
where the same sign must be chosen to preserve invariance under $\iota_{23}$.  We then obtain
\be\label{eta}
\eta = \pm\, \frac{c_0(c_1q_3^2-c_2q_2^2)}{2(c_2\sqrt{k_1}-c_1\sqrt{k_2})},
\ee
\end{subequations}
which, when substituted back into (\ref{eta2}), gives a quartic expression in the $(q_2,q_3)$ plane, with (\ref{eta}) giving an expression for $q_1^2$, thus defining a curve:
$$
(c_2q_2^2-c_1q_3^2)^2=4 (c_2\sqrt{k_1}-c_1\sqrt{k_2})(\sqrt{k_1}q_3^2-\sqrt{k_2}q_2^2),\qquad  q_1^2=q_2^2+q_3^2 \pm\, \frac{c_0(c_1q_3^2-c_2q_2^2)}{2(c_2\sqrt{k_1}-c_1\sqrt{k_2})}.
$$

\section{The Classical and Quantum Zernike System}\label{sec:zernike}

In this section, we apply the techniques of Section \ref{concurv} to the classical Zernike system \cite{17-1} and then consider the algebraic aspects of its quantisation \cite{17-2} in the framework of Section \ref{QHO}.  Of course, we mainly just reproduce formulae which can be found in \cite{17-1,17-2}, but we obtain them in a different way.  In fact, these two approaches are exactly as discussed in Section \ref{separable},  in the context of the Bertrand-Darboux equation.  Pogosyan, et al, start from separability and, as guaranteed by Jacobi's Theorem, construct constants of the motion, whilst here, we use the symmetries of the kinetic energy of the Zernike system to build commuting integrals and separable potentials.  Of course, Pogosyan, et al, do a lot more, but our purpose here is just to discuss the algebraic aspects of the super-integrability.  In the quantum case, we point out that, when written in terms of Cartesian coordinates $(x,y)$, the Zernike system is just Case IX of the Krall-Sheffer classification \cite{67-2}, so can use some of the results of \cite{f13-1}.

\subsection{The Classical Zernike System}

A classical form of the Zernike system is given in \cite{17-1}, which, in polar coordinates, takes the form
\begin{subequations}\label{zernike}
\be\label{zernike1}
H = (1 + a r^2) p_r^2+\frac{p_\phi^2}{r^2} + 2 b r p_r.
\ee
First we remove the awkward linear (in $p_r$) term by applying a gauge transformation:
$$
H = (1 + a r^2) \left(p_r+\frac{b r}{1+a r^2}\right)^2+\frac{p_\phi^2}{r^2}-\frac{b^2 r^2}{1+a r^2},
$$
so, redefining momenta by $\hat p_r = p_r+\frac{b r}{1+a r^2},\, \hat p_\phi = p_\phi$ (which are still canonical to $(r,\phi)$), and dropping ``hats'', we obtain the Zernike system in ``natural'' form:
\be\label{zernike2}
H = (1 + a r^2) p_r^2+\frac{p_\phi^2}{r^2} -\frac{b^2 r^2}{1+a r^2}.
\ee
\end{subequations}
The kinetic energy defines a constant curvature metric, with scalar curvature $R=-2a$.  This metric therefore has a 3D isometry algebra, which is easily calculated to be
\begin{subequations}\label{killing-vecs}
\be\label{killing-vecs1}
k_1 = \frac{\sqrt{1+a r^2}}{r} \left(r \cos\phi\, p_r-\sin\phi \, p_\phi\right),\quad k_2 = \frac{\sqrt{1+a r^2}}{r} \left(r \sin\phi\, p_r-\cos\phi \, p_\phi\right),\quad k_3 = p_\phi,
\ee
satisfying
\be\label{killing-vecs2}
\{k_1,k_2\} = -a k_3,\quad \{k_1,k_3\} = - k_2,\quad \{k_2,k_3\} = k_1.
\ee
\end{subequations}
The kinetic energy is just the quadratic Casimir function: $H_0 = k_1^2+k_2^2-a k_3^2$.

Adding a potential of the form $h(r)$ means that $k_3=p_\phi$ is a first integral, so the system is Liouville integrable.  In 2D we just need one additional integral for maximal superintegrability.  For example, with $H=H_0+h(r)$, we have $F_1=k_3$ and may choose a quadratic integral $F_2 = k_1k_2 + f(r,\phi)$, with $\{H,F_2\}=0$, leading to
\be\label{hf}
h(r) = \frac{b_1+b_2 r^2}{1+a r^2}, \qquad f(r,\phi) = \frac{(b_2-a b_1) r^2 \sin 2\phi}{2(1+a r^2)},
\ee
so, choosing $b_1=0, b_2=-b^2$, we obtain (\ref{zernike2}).  Whilst $\{H,F_1\}=\{H,F_2\}=0$, we must introduce a new quadratic element, defined by
$$
F_3 = \{F_1,F_2\} = k_1^2-k_2^2-\{f(r,\phi),k_3\} = k_1^2-k_2^2 + \frac{b^2r^2\cos 2\phi}{1+a r^2},
$$
where the expression for the quadratic part is derived by using the Poisson bracket relations (\ref{killing-vecs2}).  The remaining Poisson brackets are
$$
\{F_1,F_3\}=-4 F_2,\quad \{F_2,F_3\} = -F_1 (2 a H+2 a^2 F_1^2-4 b^2),
$$
and these $4$ functions satisfy the algebraic relation
$$
(H+a F_1^2)^2-F_3^2-4 F_2^2+4 b^2 F_1^2 = 0.
$$
Once again, the leading order quartic part is easily found by looking at the expressions for $F_i$ in terms of $k_i$, leaving only the ``quadratic correction'' to be determined.

\paragraph{Cartesian Coordinates $(x,y)$ and the original gauge:}  To compare results with those of \cite{17-1} we first change to Cartesian coordinate:
$$
F_1=x p_y-y p_x,\quad F_2 = (1+a (x^2+y^2)) p_xp_y-\frac{b^2 x y}{1+a (x^2+y^2)},\quad  F_3 = (1+a (x^2+y^2)) (p_y^2-p_x^2)-\frac{b^2 (y^2-x^2)}{1+a (x^2+y^2)}.
$$
Replacing the hats over the above momenta, we use the inverse gauge transformation
$$
 p_r = \hat p_r-\frac{b r}{1+a r^2},\;\;  p_\phi = \hat p_\phi \quad\Rightarrow\quad     p_x = \hat p_x-\frac{b x}{1+a (x^2+y^2)},\;\;   p_y = \hat p_y-\frac{b y}{1+a (x^2+y^2)},
$$
to obtain $F_1=x p_y-y p_x$ and
$$
 F_2 = (1+a (x^2+y^2)) p_xp_y-b (y p_x+x p_y),\quad  F_3 = (1+a (x^2+y^2)) (p_y^2-p_x^2)+2 b (x p_x- y p_y),
$$
which should be compared, respectively, with $J_1, J_3$ and $J_2$ (equations (73), (76) and (74) of \cite{17-1}).

\br
The integrals $I_2$ and $I_3$ (equations (59) and (65) of \cite{17-1}) are similarly related to $k_2^2$ and $k_1^2$.  The choice of separable coordinate corresponds, in each case, to the simultaneous diagonalisation of the pair of commuting quadratic integrals, such as $(H,F_2)$ or $(H,F_3)$ above.
\er

\subsection{The Quantum Zernike System}

For a given metric $g_{ij}$, we have the classical kinetic energy and its quantum analogue (the Laplace-Beltrami operator):
$$
H^{(2)} = \frac{1}{2} \, \sum_{i,j=1}^n g^{ij} p_ip_j \quad \mbox{and}\quad
   L_b f = \sum_{i,j=1}^n g^{ij} \nabla_i\nabla_j f = \sum_{i,j=1}^n \frac{1}{\sqrt{g}} \frac{\pa}{\pa q^j}\left(
               \sqrt{g} g^{ij}\frac{\pa f}{\pa q^i}\right),
$$
where $g$ is the determinant of the matrix $g_{ij}$.  For a metric with isometries, the
infinitesimal generators (Killing vectors) are just first order differential
operators which commute with the Laplace-Beltrami operator $L_b$.  When
the space is either flat or constant curvature, it possesses the maximal group
of isometries, which is of dimension $\frac{1}{2}n(n+1)$.  In this case, $L_b$
is proportional to the second order {\em Casimir} function of the symmetry algebra (see \cite{74-7}).

Given such a classical Hamiltonian, each Noether constant, being of degree one in momenta, generates a Hamiltonian vector field whose first $n$ components define a vector field on the configuration space, with coordinates $q_i$.  This is the corresponding Killing vector.  In the case of the classical Zernike system (\ref{zernike}), the Noether constants $k_i$ give us three Killing vectors $X_i$:
\begin{subequations}
\be\label{killv}
X_1 = \frac{\sqrt{1+a r^2}}{r} \left(r \cos\phi\, \pa_r-\sin\phi \, \pa_\phi\right),\quad X_2 = \frac{\sqrt{1+a r^2}}{r} \left(r \sin\phi\, \pa_r-\cos\phi \, \pa_\phi\right),\quad X_3 = \pa_\phi,
\ee
which, through $[X_i,X_j]=-X_{\{k_i,k_j\}}$, obey the commutation relations
\be\label{commrel}
[X_1,X_2]= a X_3,\;\;\; [X_1,X_3]= X_2,\;\;\; [X_2,X_3]=-X_1,
\ee
with
\be\label{Lb}
L_b = X_1^2+X_2^2-a X_3^2 = (1+a r^2) \pa_r^2+\left(\frac{1}{r}+2ar\right)\pa_r+\frac{1}{r^2}\, \pa_\phi^2.
\ee
Notice that the second order parts of the Laplace-Beltrami operator can be read from the metric coefficients, but that first order terms can arise (as in this case).  The {\em quantum Zernike Hamiltonian} is \cite{17-2}
\be\label{Lz}
L = L_b + (b-a) r\pa_r = (1+a r^2) \pa_r^2+\left(\frac{1}{r}+(a+b)r\right)\pa_r+\frac{1}{r^2}\, \pa_\phi^2.
\ee
\end{subequations}
We could again gauge away this additional first order term to write this as $L=L_b+h(r)$, but it is more convenient to remain in this gauge.

The construction of commuting operators is no different from finding first integrals in the classical case, except that we symmetrise our expressions where necessary:
$$
I_1 = X_3,\quad I_2 = X_1X_2+X_2X_1 +A(r,\phi) \pa_r+B(r,\phi)\pa_\phi.
$$
The condition $[L,I_2]=0$ determines the coefficients to be $A(r,\phi)=(b-a)r \sin 2\phi,\;\; B(r,\phi)=(b-a)r \cos 2\phi$.

\subsubsection{Cartesian Coordinates and Krall-Sheffer Polynomials}

Changing to Cartesian coordinates $(x,y)$, and setting $a=-1$, we find
\be\label{ks9}
L = (1-x^2) \pa_x^2-2 x y \pa_x\pa_y +(1-y^2) \pa_y^2+(b-1) (x\pa_x+y\pa_y),
\ee
which is the Krall-Sheffer operator of type IX \cite{67-2,f13-1}. Krall and Sheffer classified second order differential operators in 2D with polynomial eigenfunctions, generalising the classical results in the theory of orthogonal polynomials in 1D.  Our two commuting operators now take the form
$$
I_1 = x \pa_y - y \pa_x,\quad I_2= 2(1-x^2-y^2) \pa_x\pa_y+b (y\pa_x+x \pa_y).
$$
These generate an operator algebra which is very similar to the Poisson algebra of the classical case:
$$
\begin{array}{l}
[I_1,I_2] = 2 I_3 = 2 ((1-x^2-y^2)(\pa_x^2-\pa_y^2) + b(x \pa_x-y \pa_y)), \\[2mm]
[I_1,I_3]=-2 I_2,\quad [I_2,I_3]= 2 I_1 (2 L-2 I_1^2-b (b+2)),
\end{array}
$$
satisfying the constraint
$$
(L-I_1^2)^2 -I_2^2-I_3^2+(b^2+2b-4) I_1^2-2 (b+2) L = 0.
$$

\begin{figure}[hbt]
\begin{center}
\unitlength=0.5mm
\begin{picture}(160,80)
\put(80,80){\makebox(0,0){$P_{0 0}$}}
\put(60,60){\makebox(0,0){$P_{1 0}$}}
\put(100,60){\makebox(0,0){$P_{0 1}$}}
\put(40,40){\makebox(0,0){$P_{2 0}$}}
\put(80,40){\makebox(0,0){{$P_{1 1}$}}}
\put(120,40){\makebox(0,0){$P_{0 2}$}}
\put(20,20){\makebox(0,0){$P_{3 0}$}}
\put(60,20){\makebox(0,0){$P_{2 1}$}}
\put(100,20){\makebox(0,0){$P_{1 2}$}}
\put(140,20){\makebox(0,0){$P_{0 3}$}}
\put(0,0){\makebox(0,0){$P_{4 0}$}}
\put(40,0){\makebox(0,0){{$P_{3 1}$}}}
\put(80,0){\makebox(0,0){$P_{2 2}$}}
\put(120,0){\makebox(0,0){{$P_{1 3}$}}}
\put(160,0){\makebox(0,0){$P_{0 4}$}}
\put(75,75){\vector(-1,-1){10}}
\put(85,75){\vector(1,-1){10}}
\put(55,55){\vector(-1,-1){10}}  \put(105,55){\vector(1,-1){10}}
\put(35,35){\vector(-1,-1){10}}  \put(125,35){\vector(1,-1){10}}
\put(15,15){\vector(-1,-1){10}}  \put(145,15){\vector(1,-1){10}}
\put(80,60){\makebox(0,0){$\rightleftarrows$}}
\put(60,40){\makebox(0,0){$\rightleftarrows$}}
\put(100,40){\makebox(0,0){$\rightleftarrows$}}
\put(40,20){\makebox(0,0){$\rightleftarrows$}}
\put(80,20){\makebox(0,0){$\rightleftarrows$}}
\put(120,20){\makebox(0,0){$\rightleftarrows$}}
\put(20,0){\makebox(0,0){$\rightleftarrows$}}
\put(60,0){\makebox(0,0){$\rightleftarrows$}}
\put(100,0){\makebox(0,0){$\rightleftarrows$}}
\put(140,0){\makebox(0,0){$\rightleftarrows$}}
\put(35,55){\makebox(0,0){$R_N^+(x)$}}
\put(125,55){\makebox(0,0){$R_N^+(y)$}}
\end{picture}
\end{center}
\caption{The triangular lattice of polynomials $P_{m n}$
with $P_{0 0}=1$.  Horizontal arrows denote the
action of $J_m^+$ (right) and $J_n^-$ (left). The operators $R_N^+(x)$ and $R_N^+(y)$ represent the edge ladder operators.} \label{trilatt}
\end{figure}
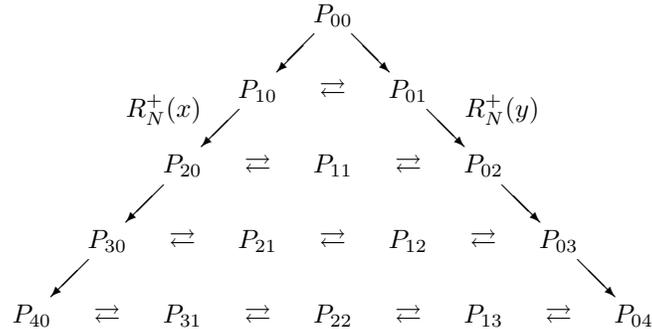

Consider {\em monic} polynomial eigenfunctions of $L$:
\be  \label{pmn}  %
P_{mn} = x^m y^n + \mbox{lower order terms} ,\quad\mbox{for}\;\;\; m+n=N .
\ee  %
The eigenvalue for $P_{mn}$ is easily calculated from the leading order term to be $\lambda_{m+n}=(m+n) (b-m-n)$, with the notation reflecting the fact that it is the {\em same} eigenvalue for all $(m,n)$ with $m+n=N$.  This means that in Figure \ref{trilatt}, all polynomials on a horizontal line have the same eigenvalue.  This degeneracy reflects the existence of the commuting operators, which do not change the eigenvalue.

If we act on $P_{mn}$ with $I_1$ and $I_2$, then, looking at the leading order terms, we find
$$
I_1P_{mn} = nP_{m+1\,n-1}-mP_{m-1\, n+1},\quad I_2P_{mn} = n(b-2m)P_{m+1\, n-1}+m(b-2n) P_{m-1\, n+1},
$$
so the operators
\begin{subequations}
\be\label{jpm}
\begin{array}{l}
J_m^+ = I_2-(b-2 m) I_1 = 2 (m x \pa_y+(b-m) y \pa_x+(1-x^2-y^2)\pa_x\pa_y),\\[2mm]
 J_n^- = I_2+(b-2 n) I_1 = 2 (n y \pa_x+(b-n) x \pa_y+(1-x^2-y^2)\pa_x\pa_y),
 \end{array}
\ee
move us to right or left across the horizontal levels:
\be\label{rightleft}
J_m^+P_{mn}=2m(b-m-n)P_{m-1\, n+1},\qquad J_n^-P_{mn}=2n(b-m-n)P_{m+1\, n-1}.
\ee
\end{subequations}

Notice that any eigenfunction of (\ref{ks9}), which is independent of $y$, just satisfies the 1D Gegenbauer equation,
$$
L_xP_{N 0} = ((1-x^2) \pa_x^2+(b-1) x\pa_x)P_{N 0} =\lambda_N P_{N 0},\quad\mbox{with}\quad  \lambda_N = N (b-N),
$$
for which we have a standard raising operator:
$$
R_N^+ = \frac{1}{b-2N} ((1-x^2) \pa_x +(b-N) x),\quad\mbox{with}\quad [L_x,R_N^+]=\frac{2x}{2N-b} (L_x-\lambda_N) + (\lambda_{N+1}-\lambda_N)R_N^+,
$$
generating Gegenbauer polynomials on the left edge of Figure \ref{trilatt}, with $J_m^+$ taking us to the polynomials $P_{mn}$, for which $m+n=N$.  Notice that $J_0^+P_{0 N}=0$, since $m=0$.

\br
The operator (\ref{ks9}) is symmetric in $x$ and $y$, so the polynomials on the right edge of Figure \ref{trilatt} are also Gegenbauer polynomials, but this time, in the variable $y$.  We could, therefore, have started on the right edge and used $J_n^-$ to move to the left, with $J_0^-P_{N 0}=0$.
\er

\section{Conclusions}

Maximally super integrable systems are distinguished by their number of first integrals or commuting operators.  In this paper, the emphasis has been on the {\em role} of these in constructing trajectories or eigenfunctions. In the classical case, the level set of the $2n-1$ first integrals is a curve, representing the unparameterised trajectory.  In the quantum case, the existence of commuting operators is responsible for the degeneracy of eigenvalues.  The additional eigenfunctions are connected through the action of the commuting operators.  The emphasis throughout has been that symmetries of the kinetic energy have a major influence on the structure of completely/super-integrable potentials and their corresponding Poisson algebras.  As seen above, the Poisson algebras in 2D are very simple, but in higher dimensions (even in 3D), can be very complicated.

The term ``quantum integrable'' is not only used to describe {\em physical, quantum} systems, but also any (large enough) family of commuting differential operators. In particular, there is a theory of orthogonal polynomials in $2-$dimensions (and higher) and an important subclass of these are {\em eigenfunctions} of some sort of differential operator.

In \cite{67-2}, Krall and Sheffer classified second order, linear differential operators in 2D with polynomial eigenfunctions. These Krall-Sheffer polynomials are 2D generalisations of the classical orthogonal polynomials, so much of the standard machinery can be derived.  In \cite{f13-1} we construct $3-$level recurrence relations (depending upon at most $9$ points) for each of the $9$ cases, as well as raising operators in {\em both} directions (parallel to, but not on, the edges), and discuss the reduction of these to $1-$dimensional cases on the edges.  In particular, for case IX (the Zernike case) we construct $4-$point, $3-$level recurrence relations, raising operators in {\em both} directions, as well as a {\em generating function}.


\begin{thebibliography}{10}

\bibitem{f05-1}
A.P. Fordy.
\newblock Symmetries, ladder operators and quantum integrable systems.
\newblock {\em Glasgow Mathematical Journal}, 47A:65--75, 2005.

\bibitem{f06-1}
A.P. Fordy.
\newblock Darboux related quantum integrable systems on a constant curvature
  surface.
\newblock {\em J.Geom. and Phys.}, 56:1709--27, 2006.

\bibitem{f07-1}
A.P. Fordy.
\newblock Quantum super-integrable systems as exactly solvable models.
\newblock {\em SIGMA}, 3:025, 10 pages, 2007.
\newblock http://dx.doi.org/10.3842/SIGMA.2007.025.

\bibitem{13-2}
W.~Miller Jr, S.~Post, and P.~Winternitz.
\newblock Classical and quantum superintegrability with applications.
\newblock {\em J.Phys.A}, 46:423001 (97 pages), 2013.

\bibitem{14-2}
M.~Cariglia.
\newblock Hidden symmetries of dynamics in classical and quantum physics.
\newblock {\em Rev. Mod. Phys.}, 86:1283--1333, 2014.

\bibitem{f17-5}
A.P. Fordy and Q.~Huang.
\newblock {Poisson} algebras and 3d superintegrable {Hamiltonian} systems.
\newblock 2017.
\newblock preprint arXiv:1708.07024 [nlin.SI].

\bibitem{17-1}
G.S. Pogosyan, K.B. Wolf, and A.~Yakhno.
\newblock Superintegrable classical {Zernike} system.
\newblock {\em J.Math.Phys.}, 58:072901, 2017.

\bibitem{17-2}
G.S. Pogosyan, C.~Salto-Alegre, K.B. Wolf, and A.~Yakhno.
\newblock Quantum superintegrable {Zernike} system.
\newblock 2017.
\newblock arXiv:1702.08570 [math-ph].

\bibitem{67-2}
H.L. Krall and I.M. Sheffer.
\newblock Orthogonal polynomials in two variables.
\newblock {\em Ann.Mat.Pura Appl. ser. 4}, 76:325--76, 1967.

\bibitem{f13-1}
A.P. Fordy and M.J. Scott.
\newblock Recursive procedures for {Krall-Sheffer} operators.
\newblock {\em J.Math.Phys}, 54:043516 (23 pages), 2013.
\newblock http://arxiv.org/abs/1211.3075.

\bibitem{40-2}
J.M. Jauch and E.L. Hill.
\newblock On the problem of degeneracy in quantum mechanics.
\newblock {\em Phys.Rev}, 57:641--5, 1940.

\bibitem{74-7}
R.~Gilmore.
\newblock {\em Lie Groups, Lie Algebras and Some of Their Applications}.
\newblock Wiley, New York, 1974.

\bibitem{76-8}
L.D. Landau and E.M. Lifshitz.
\newblock {\em Course of Theoretical Physics Volume 1: Mechanics}.
\newblock Pergamon, Oxford, 1976.

\bibitem{88-4}
E.T. Whittaker.
\newblock {\em A Treatise on the Analytical Dynamics of Particles and Rigid
  Bodies}.
\newblock CUP, Cambridge, UK, 1988.

\bibitem{88-8}
I.~Marshall and S.~Wojciechowski.
\newblock When is a {Hamiltonian} system separable?
\newblock {\em J.Math.Phys.}, 29:1338--46, 1988.

\bibitem{08-7}
R.G. Smirnov.
\newblock The classical {Bertrand–-Darboux} problem.
\newblock {\em J.Math.Sci.}, 151:3230--44, 2008.

\end{thebibliography}

\end{document}